\documentclass[conference,10pt,table]{IEEEtran}

\usepackage[utf8]{inputenc}
\usepackage{array} 
\usepackage{epsfig}
\usepackage{epstopdf}
\usepackage{times}
\usepackage{framed}
\usepackage{enumitem}
\usepackage{float}
\usepackage{etoolbox}
\usepackage{algpseudocode}
\usepackage{amssymb}
\usepackage{amsmath}
\usepackage{multirow}
\usepackage{url}
\usepackage{lipsum}
\usepackage{soul}
\usepackage{caption}
\usepackage{mathtools}
\usepackage{bbm}
\usepackage{booktabs}
\usepackage{xcolor}
\usepackage{makecell}
\usepackage{pifont}
\usepackage{subfigure}
\usepackage[export]{adjustbox} 
\usepackage{latexsym}
\usepackage{cite}
\usepackage[acronym]{glossaries}
\usepackage{multicol}
\usepackage{eurosym}
\usepackage{xspace}
\usepackage{pdfpages}
\usepackage{verbatim}
\usepackage{stfloats}
\usepackage{mathtools}
\usepackage[norelsize,boxruled]{algorithm2e}

\newcommand{\vv}[1]{\mathbf{#1}}

\newcommand{\Compl}{\mathbb{C}}

\newcommand{\herm}{\mathrm{H}}
\newcommand{\tr}{\mathrm{tr}}
\newcommand{\tran}{\mathrm{T}}

\newcommand{\alphav}{\boldsymbol{\alpha}}
\newcommand{\Real}{\mbox{$\mathbb{R}$}}

\newcommand{\name}{FABRIS}

\newacronym{ula}{ULA}{uniform linear array}
\newacronym{csit}{CSIT}{channel state information at the transmitter}
\newacronym{cdf}{CDF}{cumulative distribution function}
\newacronym[plural=BSs, firstplural=base stations (BSs)]{bs}{BS}{base station}
\newacronym{pdf}{pdf}{probability distribution function}
\newacronym{aod}{AoD}{angle of departure}
\newacronym{aoa}{AoA}{angle of arrival}
\newacronym{ue}{UE}{user equipment}
\newacronym{los}{LoS}{line-of-sight}
\newacronym{pla}{PLA}{planar linear array}
\newacronym[plural=RISs, firstplural=reconfigurable intelligent surfaces (RISs)]{ris}{RIS}{reconfigurable intelligent surface}
\newacronym{sdp}{SDP}{semidefinite programming}
\newacronym{sdr}{SDR}{semidefinite relaxation}
\newacronym{slnr}{SLNR}{signal-to-leakage-and-noise ratio}
\newacronym{sre}{SRE}{smart radio environment}
\newacronym{snr}{SNR}{signal-to-noise ratio}
\newacronym{toa}{ToA}{time-of-arrival}
\newacronym{doa}{DoA}{direction-of-arrival}
\newacronym{mmse}{MMSE}{minimum mean squared error}
\newacronym{peb}{PEB}{position error bound}
\newacronym{oeb}{OEB}{orientation error bound}
\newacronym{rss}{RSS}{received signal strength}
\newacronym{ml}{ML}{machine learning}
\newacronym{rmse}{RMSE}{root-mean-square error}
\newacronym{mmwave}{mm-Wave}{millimeter-wave}
\newacronym{csi}{CSI}{channel state information}
\newacronym{3gpp}{3GPP}{3rd Generation Partnership Project}
\newacronym{sinr}{SINR}{signal-to-interference-plus-noise ratio}
\newacronym{sbr}{SBR}{shooting and bouncing rays}
\newacronym{ura}{URA}{uniform rectangular array}
\newacronym{ofdma}{OFDMA}{orthogonal frequency-division multiple access}
\newacronym[plural=CSs, firstplural=candidate sites (CSs)]{cs}{CS}{candidate site}
\newacronym{mrt}{MRT}{maximum ratio transmission}
\newacronym{bca}{BCA}{Block Coordinate Ascent}
\newacronym{fp}{FP}{Fractional Programming}
\newacronym{tdma}{TDMA}{time-division multiple access}
\newacronym{jfi}{JFI}{Jain's fairness index}
\newacronym[plural=UAVs, firstplural=unmanned aerial vehicles (UAVs)]{uav}{UAV}{unmanned aerial vehicle}
\newacronym{ios}{IoS}{Internet of Surfaces}
\newacronym{soa}{SoA}{state-of-the-art}
\newacronym[plural=RFs, firstplural=radio frequencies (RFs)]{rf}{RF}{radio frequency}
\newacronym[plural=RATs, firstplural=radio access technologies (RATs)]{rat}{RAT}{radio access technologies}
\newacronym{pin}{PIN}{positive-intrinsic-negative}
\newacronym{em}{EM}{electromagnetic}
\newacronym{pec}{PEC}{perfect electric
conductor}
\newacronym{nema}{NEMA}{National Electrical Manufacturers Association}

\newcommand{\rmris}{\textnormal{\tiny{RIS}}}
\newcommand{\rmd}{\textnormal{\tiny{d}}}
\newcommand{\rmue}{\textnormal{\tiny{UE}}}
\newcommand{\rmbs}{\textnormal{\tiny{BS}}}

\IEEEoverridecommandlockouts

\begin{document}

\setlength{\textfloatsep}{3pt}

\title{A Frequency-Agnostic RIS-based solution to control the Smart Radio Propagation Environment}

\author{
     \IEEEauthorblockN{Fabio Maresca\IEEEauthorrefmark{1}\IEEEauthorrefmark{2}, Antonio Albanese\IEEEauthorrefmark{1}\IEEEauthorrefmark{3}, Placido Mursia\IEEEauthorrefmark{1},
     Vincenzo Sciancalepore\IEEEauthorrefmark{1}, Xavier Costa-P\'erez\IEEEauthorrefmark{4}\IEEEauthorrefmark{1}}
     \IEEEauthorblockA{
 	\IEEEauthorrefmark{1}NEC Laboratories Europe, 69115 Heidelberg, Germany\\
  	\IEEEauthorrefmark{2}Department of Network Engineering, Polytechnic University of Catalonia, 08034 Barcelona, Spain\\
  	\IEEEauthorrefmark{3}Department of Telematic Engineering, University Carlos III of Madrid, 28911 Legan\'es, Spain\\
  	\IEEEauthorrefmark{4}i2cat Foundation and ICREA, 08034 Barcelona, Spain
  	\\\{name.surname\}@neclab.eu}
  	\thanks{This work was supported by EU H2020 RISE-6G (grant agreement 101017011) and EU H2020 SMARTFACT (grant agreement 956670) projects.}
 }

\maketitle

\begin{abstract}

The disruptive \gls{ris} technology is steadily gaining relevance as a key element in future 6G networks. 
However, a one-size-fits-all \gls{ris} hardware design is yet to be defined due to many practical considerations. A major roadblock for currently available \glspl{ris} is their inability to concurrently operate at multiple carrier frequencies, which would lead to redundant installations to support multiple \glspl{rat}. 

In this paper, we introduce \name{}, a novel and practical multi-frequency \gls{ris} design. \name{} is able to dynamically operate across different \glspl{rf} by means of frequency-tunable antennas as unit cells with virtually no performance degradation when conventional approaches to \gls{ris} design and optimization fail. Remarkably, our design preserves a sufficiently narrow beamwidth as to avoid generating signal leakage in unwanted directions and a sufficiently high antenna efficiency in terms of scattering parameters. Indeed, \name{} selects the \gls{ris} configuration that maximizes the signal at the intended target \gls{ue} while minimizing leakage to non-intended neighboring \glspl{ue}. Numerical results and full-wave simulations validate our proposed approach against a naive implementation that does not consider signal leakage resulting from multi-frequency antenna arrays.
\end{abstract}

\glsresetall

\section{Introduction}

\Glspl{ris} have the potential to drive the biggest paradigm change in classical wireless network design. By dynamically altering the propagation of the impinging signals, they transmute the nature of the wireless radio channel from the long-established tamper-proof black-box to a variable that can be optimized, thus sowing the seeds of a fully controllable and flexible smart propagation environment~\cite{direnzo2020}. \glspl{ris} draw modest power and have limited installation and maintenance costs, being envisioned as planar arrays of cheap ultra-thin reflective elements capable of generating distinct (in practice, discrete) phase shifts according to the desired \gls{ris} beamforming configuration~\cite{Dai2020,Mursia2020}. \glspl{ris} are thus key-enablers for ubiquitous network deployments: mounted on walls (akin to wallpapers), building facades or billboards, they may boost network throughput and localization performance while providing emergency services in disaster scenarios when boarded on \glspl{uav}~\cite{albanese2021}. Although most \gls{soa} solutions assume a reliable \gls{ris} control channel to the on-board \gls{ris} controller, recent developments in \gls{ios} have shed light on autonomous \glspl{ris} without a dedicated control channel, thereby further improving their deployment flexibility~\cite{albanese2022}.

However, \glspl{ris} practicality is threatened by rigid hardware implementations. Conventional designs support a single operating band, which does not match the diverse and discontinuous mobile \gls{rf} spectrum allocations. In this context, network operators would not be able to adjust the \glspl{ris} operating band as needed.  This calls for redundant deployments with multiple \glspl{ris}, each of them designed to operate on a specific band-of-interest, thus increasing the associated CAPEX and OPEX incurred while deploying \gls{ris}-enabled networks~\cite{RIScommag_2021}.

\textbf{Related Work.} \gls{ris} design choices in terms of array technology, total number of elements, and inter-element spacing directly affect the achievable performance of \gls{ris}-aided wireless networks~\cite{Trichopoulos2021,Gros2021}. In particular, the operating frequency depends on the reflective element geometry and material, whereas the total number of elements $N$ determines the beamforming gain, which in turn may increase the \gls{snr} at the receiver up to a factor proportional to $N^2$~\cite{Qingqing2021}. A common arrangement in the current \gls{soa} is to set the inter-element distance to half the operating wavelength since different values might lead to unwanted phenomena, negatively affecting the overall system performance if not taken into account~\cite{Dunna2020,Pei2021}: $i$) smaller values result in \emph{mutual coupling} across the array, correlating the transmitted signals at different antennas; $ii$) considerably larger values generate \emph{grating lobes}, i.e., secondary and uncontrollable lobes in the array response that spread energy in unwanted directions and may ultimately create interference among receivers. It is important to remark that, to the best of our knowledge, in all such existing \gls{ris} designs the operating frequency is fixed by design.

\textbf{Contributions.} Inspired by frequency-reconfigurable antennas (see, e.g., \cite{Aftab2015}), we propose to endow each \gls{ris} element with the capability to programmatically switch the operating frequency in real-time.\footnote{Note that the associated control signalling is enabled by the existing protocols conventionally used to reconfigure the phase-shifting configuration of each \gls{ris} element.} However, as a result, the inter-element distance becomes in general different than the conventional half-wavelength, leading to potentially negative effects on the performance as described above. To counteract this issue, we equip each RIS element with a matched load that can be activated to dissipate the incoming signal and effectively reduce the number of reflecting \gls{ris} elements. Such mechanism allows increasing the effective inter-element spacing, which reduces the mutual coupling across the array at the expense of a lower number of active \gls{ris} elements. In some relevant applications, such as multicasting, this approach may be further exploited to enhance the presence of grating lobes and turn them into an advantage, thereby spreading the signal across a wide angular span. Since it is not possible to reduce the inter-element spacing below the fabrication value, we select the latter to obtain a half-wavelength spacing for the case of minimum value of signal wavelength (i.e., the maximum operating frequency). More specifically, in this paper, $i$) we design a novel \gls{ris} reflective element as a multi-frequency patch antenna whose working frequency can be dynamically set to $21.28$ GHz or $27.96$ GHz, $ii$) we develop a \gls{ris} array design that leverages on suitably matched loads to disable the reflections at specific elements and control the phase shift for both working frequencies, $iii$) validate the corresponding \gls{ris} antenna array design by means of full-wave simulations, $iv$) formulate an optimization problem targeting the maximization of the \gls{slnr} at the target \gls{ue} by means of suitably choosing the \gls{ris} phase shifts configuration and elements activation, $v$) show that our algorithm outperforms the naive approach of activating all the antenna elements, i.e., not controlling the effective inter-element distance. To the best of our knowledge, our solution, denoted as Frequency-Agnostic Behavior \gls{ris} (\name{}), is the \emph{first work} presenting a practical and efficient approach to operate a \gls{ris} at multiple carrier frequencies, enabling frequency selection at run-time.


\textbf{Notation}. We denote matrices and vectors in bold while each of their element is indicated in roman with a subscript. $(\cdot)^{\tran}$ and $(\cdot)^{\herm}$ stand for vector or matrix transposition and Hermitian transposition, respectively. The L$2$-norm of a vector and the trace of a square matrix are denoted by $\| \cdot \|$ and $\tr(\cdot)$, respectively.

\addtolength{\topmargin}{-0.14in}
\section{System model}
\label{s:system_model}

\addtolength{\topmargin}{+0.28in}

We consider a network comprised of a single-antenna \gls{bs}, a \gls{ris} equipped with $N = N_x N_y$ antenna elements, where $N_x$ ($N_y$) is the number of antenna elements along the $x$ ($y$)-axis, and a single-antenna \gls{ue}.\footnote{In this context, we consider the simple single-antenna case in order to better highlight the characteristics of our proposed approach. However, note that our approach can be readily extended to the case of multiple antennas at the \gls{bs} and at the \gls{ue}, which is thus left for future work.} Without loss of generality, we assume the \gls{bs} to be placed at the origin of our reference system, whereas the \gls{ris} is placed at $\vv{p}_{\rmris}\in\Real^{3\times 1}$, and the \gls{ue} at $\vv{p}_{\rmue}\in\Real^{3\times 1}$. In this paper, we study the effect of varying the working frequency of the available \gls{ris} on the network performance. Hence, in the following we will express the relevant channel vectors as a function of the working wavelength $\lambda$. Let $\vv{a}(\theta,\phi,\lambda) \in \Compl^{N\times 1}$ be the \gls{pla} response at the \gls{ris} for the steering angles $(\theta,\phi)$ along the azimuth and elevation, respectively, be defined as
\begin{align}\label{eq:PLA}
    \vv{a}(\theta,\phi,\lambda) &\!\triangleq\! [1, e^{-j2\pi\frac{d}{\lambda}\cos(\theta)}, \ldots, e^{j-2\pi\frac{d}{\lambda}(N_x-1)\cos(\theta)}]^\tran \nonumber \\
    \phantom{=} & \!\otimes\! [1, e^{-j2\pi\frac{d}{\lambda}\sin(\phi)}, \ldots, e^{-j2\pi\frac{d}{\lambda}(N_y-1)\sin(\phi)}]^\tran.
\end{align}
Assuming \gls{los} propagation, we let $\vv{h}\in\Compl^{N\times1}$ denote the channel from the \gls{ris} to the \gls{ue}, which is defined as $\vv{h} \triangleq \sqrt{\gamma_{\rmue}} \,\, \vv{a}(\theta{\rmue},\phi{\rmue},\lambda)$, with $\gamma_{\rmue} \triangleq \frac{\beta_0}{\|\vv{p}_{\rmue}-\vv{p}_{\rmris}\|^2}$ the average channel power gain, and $\beta_0$ the average channel power at a reference distance. In a similar way, we define the channel from the \gls{bs} to the \gls{ris} and the direct link from the \gls{bs} to the \gls{ue} as
\begin{align}
    \vv{g} &  \triangleq \sqrt{\gamma_{\rmbs}} \,\, \vv{a}(\theta_{\rmbs},\phi_{\rmbs},\lambda)\\
    h_{\rmd} & \triangleq \sqrt{\gamma_{\rmd}} e^{\frac{j2\pi}{\lambda}\|\vv{p}_{\rmue}\|},
\end{align}
with $\gamma_{\rmbs} \triangleq \frac{\beta_0}{\|\vv{p}_{\rmris}\|^2}$, and $\gamma_{\rmd} \triangleq \frac{\beta_0}{\|\vv{p}_{\rmue}\|^2}$, respectively. 

We remark that in Eq.~\eqref{eq:PLA}, the inter-element distance $d$ is fixed during the fabrication phase, whereas the working wavelength $\lambda$ is a system parameter. Hence, in general $d\neq \lambda/2$, leading to potential mutual coupling or grating lobes across the array for the case of $d<\lambda/2$ and $d\gg\lambda/2$, respectively. To counteract such issue, our envisioned multi-frequency \gls{ris} possesses the unique ability to turn-off its elements, thereby dynamically reconfiguring the total number of effective \gls{ris} elements and the effective inter-element spacing. Such capability is modelled by the \emph{activation profile} $\alphav \in \{0,1\}^{N\times 1}$ and the corresponding \emph{activation matrix} $\vv{A}(\alphav)= \mathrm{diag}(\alphav)$, where $\alpha_n = 0$ indicates that the $n$-th \gls{ris} element is turned-off, and vice-versa. We thus define the receive signal at the \gls{ue} as
\begin{align}
    y \triangleq (\vv{h}^\herm\vv{A}(\alphav)\vv{\Phi}\vv{g} + h_{\rmd}) s + n \in \Compl
\end{align}
with $\vv{\Phi} \triangleq \mathrm{diag}(e^{j\phi_1},\ldots,e^{j\phi_N})\in\Compl^{N\times N}$ and $\phi_n \in [0,2\pi), \quad \forall n$, $s\in \Compl$ the transmit signal at power $P$ with $\|s\|^2 = P$ and $n\sim\mathcal{CN}(0,\sigma_n^2)$ the noise term.

\section{Problem Formulation}\label{s:problem}

Our objective is to jointly optimize the \gls{ris} phase shifts and activation profiles $\vv{\Phi}$ and $\vv{A}(\alphav)$, respectively, by maximizing the \gls{slnr} of the intended \gls{ue} location. Let $T$ be the number of non-intended \glspl{ue} distributed across a given target area, which are in positions $\{\vv{p}_t\}_{t=1}^T$ and have corresponding channels given by $\{\vv{h}_t\}_{t=1}^T$ and $\{h_{\rmd,t}\}_{t=1}^T$. Hence, we define the \gls{slnr} as
\begin{align}
    \mathrm{SLNR}(\vv{\Phi},\alphav,\lambda) \triangleq \frac{|(\vv{h}^\herm\vv{A}(\alphav)\vv{\Phi}\vv{g} + h_{\rmd})|^2}{\frac{\sigma_n^2}{P} +  \sum_{t=1}^T |(\vv{h}_t^\herm\vv{A}(\alphav)\vv{\Phi}\vv{g} + h_{\rmd,t})|^2}.
\end{align}

Our objective is thus formalized as follows
\begin{align}
    \begin{array}{cl}
        \displaystyle \max_{\vv{\Phi}, \alphav} & \mathrm{SLNR}(\vv{\Phi},\alphav,\lambda) \\
         \mathrm{s.t.} &  |\Phi_{ii}|^2 = 1, \quad |\Phi_{ij}|^2 = 0 \quad \forall i\neq j \\
         & \alpha_i \in \{0,1\} \quad \forall i,
    \end{array}
    \label{problem:max_SLRN}
\end{align}
which is particularly complex to tackle due to its non-convex formulation and the dependency of the two optimization variables. Therefore, in the following we solve problem \eqref{problem:max_SLRN} by decoupling the optimization in two separate sub-problems via alternating optimization.

\section{Convex relaxation and solution}
\label{s:algorithm}

In this section, we provide a solution for the optimization problem in~\eqref{problem:max_SLRN}. Specifically, in addition to the \gls{ris} phase shifts $\vv{\Phi}$, we derive the activation profile $\vv{A}$, which is fundamental for our \name{} system design.    
Since there is a single \gls{ue} in the system and assuming perfect \gls{csit}, we can safely set the \gls{ris} phase profile to maximize the signal power at its location. Let $\vv{v} = \mathrm{diag}(\vv{\Phi}^\herm)$ such that the \gls{ris} configuration is set to the following
\begin{align}
    \vv{v} = \mathrm{exp}\left[j (\angle\bar{\vv{h}}  - \angle h_{\rmd} )\right]
\end{align}
with $\bar{\vv{h}} = \mathrm{diag}(\vv{h}^\herm)\vv{g}$. We are now left with the problem of optimizing the \gls{ris} activation profile, which is formalized via the following binary optimization problem
\begin{align}\label{eq:Prob_binary}
    \begin{array}{cl}
        \displaystyle \max_{\bar{\alphav}} & \frac{\bar{\alphav}^\tran\vv{H}\vv{H}^\herm\bar{\alphav}}{\bar{\alphav}^\tran\vv{G}\vv{G}^\herm\bar{\alphav} + \sigma_n^2/P} \\
         \mathrm{s.t.} &  \bar{\alpha}_i \in \{0,1\} \quad i=1,\ldots,N,\\
         & \bar{\alpha}_{N+1} = 1
    \end{array}
\end{align}
where we have defined $\bar{\alphav} = \begin{bmatrix}\alphav^\tran & 1\end{bmatrix}^\tran\in\Compl^{N+1\times 1}$ $\vv{H} = \begin{bmatrix}(\mathrm{diag}(\vv{h}^\herm)\vv{\Phi}\vv{g})^\tran & h_{\rmd}\end{bmatrix}^\tran\in \Compl^{N+1\times 1}$ and $\vv{G} = \sum_{t}\begin{bmatrix}(\mathrm{diag}(\vv{h}_t^\herm)\vv{\Phi}\vv{g})^\tran & h_{\rmd,t}\end{bmatrix}^\tran\in \Compl^{N+1\times 1}$. We tackle Problem \eqref{eq:Prob_binary} by firstly relaxing the non-convex binary constraint to $0\leq \bar{\alpha}_i \leq 1, \quad i=1,\ldots,N$ and defining the positive semi-definite matrix variable $\vv{V} = \bar{\alphav}\bar{\alphav}^\tran$ with the associated constrain on its rank, i.e., $\mathrm{rank}(\vv{V}) = 1$. Next, we employ \gls{sdr} and the bisection method to recover a solution $\vv{V}^\star$. Finally, we extract a rank-$1$ solution $\alphav^\star$ by employing Gaussian randomization and quantizing the resulting candidate vectors onto the binary set. Note that the aim of such optimization routine is to reshape the effective antenna array response at the \gls{ris} so as to minimize leakage at neighboring \glspl{ue} while preserving the \gls{snr} at the target \gls{ue}.

\section{Design of multi-frequency RIS}\label{s:design}

\begin{figure}[t!]
    \centering  
    \includegraphics[clip,width =  0.85\linewidth]{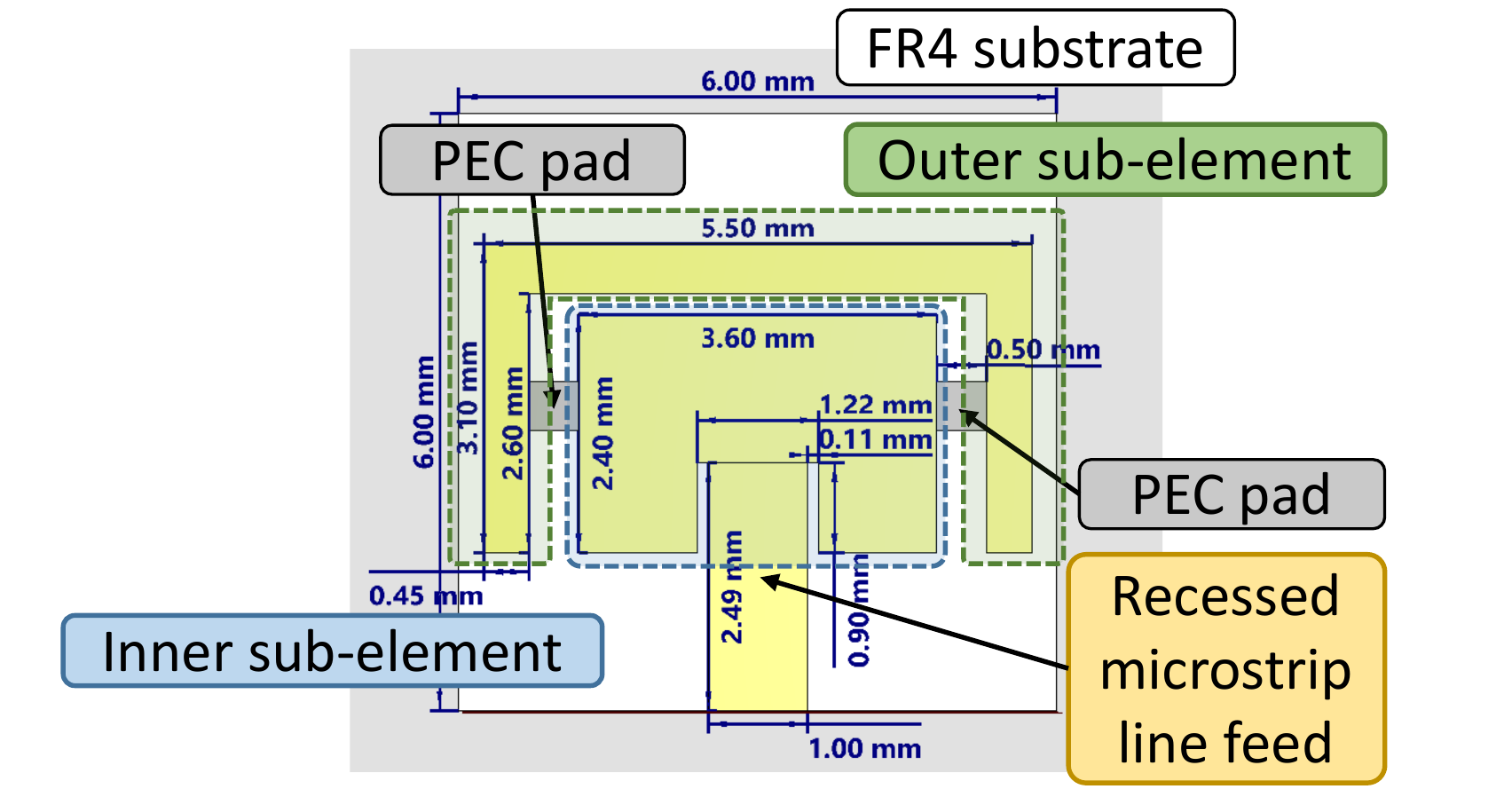}
    \caption{CST model of one \gls{ris} reflective element consisting of two patch antennas connected by \gls{pec} pads.}
    \label{fig:patch_antenna}
\end{figure}

Our solution, namely \name{}, complements the \gls{ris} optimization described above with an ad-hoc hardware design based on modular patch antennas capable of supporting multiple frequencies in a reconfigurable fashion.
Specifically, since the main radiating frequency of an antenna depends on its physical geometry, we envision to divide each \gls{ris} element into two sub-elements, as shown in Fig.~\ref{fig:patch_antenna}. By activating only the inner sub-element, the \gls{ris} operates at $f_1 = 27.96$ GHz while working at $f_2 = 21.28$ GHz when both sub-elements are active.  

\textbf{Frequency-agnostic operation.} The electrical link between the sub-elements is a \gls{pin} diode, which under zero- or reverse-bias (\textsc{off} state) attenuates most of the \gls{rf} signal, thus isolating the two sub-elements, while behaving as a good \gls{rf} conductor under forward bias (\textsc{on} state). Moreover, \gls{pin} diodes are often employed in \gls{rf} circuits due to their fast switching times (in the order of hundreds of nanoseconds), thereby allowing us to set the \gls{ris} working frequency at run-time with no service disruption. Although \gls{ris} design often borrows from antenna theory, \glspl{ris} are intrinsically different from radiating antennas as they do not transmit locally generated \gls{rf} signals but only reflect the incoming \gls{em} waves. Hence, setting \textsc{off} the \gls{pin} diode does not prevent the reflection of the incoming \gls{em} energy onto both sub-elements. Therefore, whenever the \gls{pin} diode is \textsc{off}, we envision to connect the outer sub-element to a load whose impedance is matched to the characteristic impedance of such sub-element. As a result, the \gls{em} power impinging on it is absorbed, thus eliminating unwanted reflections while operating the \gls{ris} at frequency $f_1$. 

In order to introduce variable yet discrete phase shifts at each \gls{ris} element, we employ a set of microstrip-lines, whose length $l$ is given by $l = \frac{\phi}{2} \frac{v_f \, c}{2 \pi f}$~\cite{balanis2012}, where $\phi$ is the desired phase shift, $v_f$ is the velocity factor of the microstrip-line, $c$ is the speed of light in vacuum and $f$ is the working frequency. As the microstrip-line length depends on the selected frequency, two transmission lines are needed to provide a single phase shift at the two operating frequencies, i.e. $f_1$ and $f_2$. Hence, the \gls{ris} controller makes use of two \gls{rf} switches per \gls{ris} element, each of them dedicated to a single sub-element, to route the \gls{rf} signal to a specific microstrip-line in light of the required phase shift on the reflected signal. Besides, the \gls{rf} switch downstream of the outer sub-element has an additional port for the matched load to support full power absorption when operating the \gls{ris} at $f_1$, i.e. with \gls{pin} diode \textsc{off}. It is worth highlighting that further engineering of the working frequencies may reduce the number of required transmission lines at the expense of a more complex switching network. For instance, if $f_1 = k f_2$, with $k > 0$, the same transmission line introducing a phase shift $\phi$ at $f_1$ provides a phase shift $k \phi$ at $f_2$ (net of phase wrapping around $2 \pi$). 

\begin{table}[t]
\caption{CST design parameters}
\label{tab:CST_parameters}
\centering
\resizebox{\linewidth}{!}{%
\begin{tabular}{cc|cc}
\textbf{Parameter} & \textbf{Value} & \textbf{Parameter} & \textbf{Value}\\  
\hline
\rowcolor[HTML]{EFEFEF}
Dielectric thickness             & $0.36$ mm                 & Dielectric material           & FR-4 (lossy)       \\
Antenna material    & copper (annealed)    & Antenna thickness  & $0.035$ mm        \\

\end{tabular}%
}
\end{table}

\textbf{Full-wave simulation.} Let us validate our \gls{ris} design by means of full-wave simulations in CST Studio Suite 2021~\cite{CST}. The geometry of a single \gls{ris} element is shown in Fig.~\ref{fig:patch_antenna}, wherein the circuital behavior of the connecting \gls{pin} diode is modeled by means of \gls{pec} pads~\cite{Kishk_13}. In particular, when the diode is \textsc{ON}, the sub-elements are physically connected through two \gls{pec} pads, whereas no pad is used with the diode in the \textsc{off} state. The physical dimensions of the \gls{pec} pads reflect the typical size of a \gls{pin} diode, see e.g.~\cite{pin}. Moreover, the patch antenna is matched using a recessed microstrip-line feed while we make use of \gls{nema} FR4 as substrate material. The design and simulation parameters are listed in Table~\ref{tab:CST_parameters}.

\begin{figure}[t!]
    \centering  
    \includegraphics[clip,width =  \linewidth]{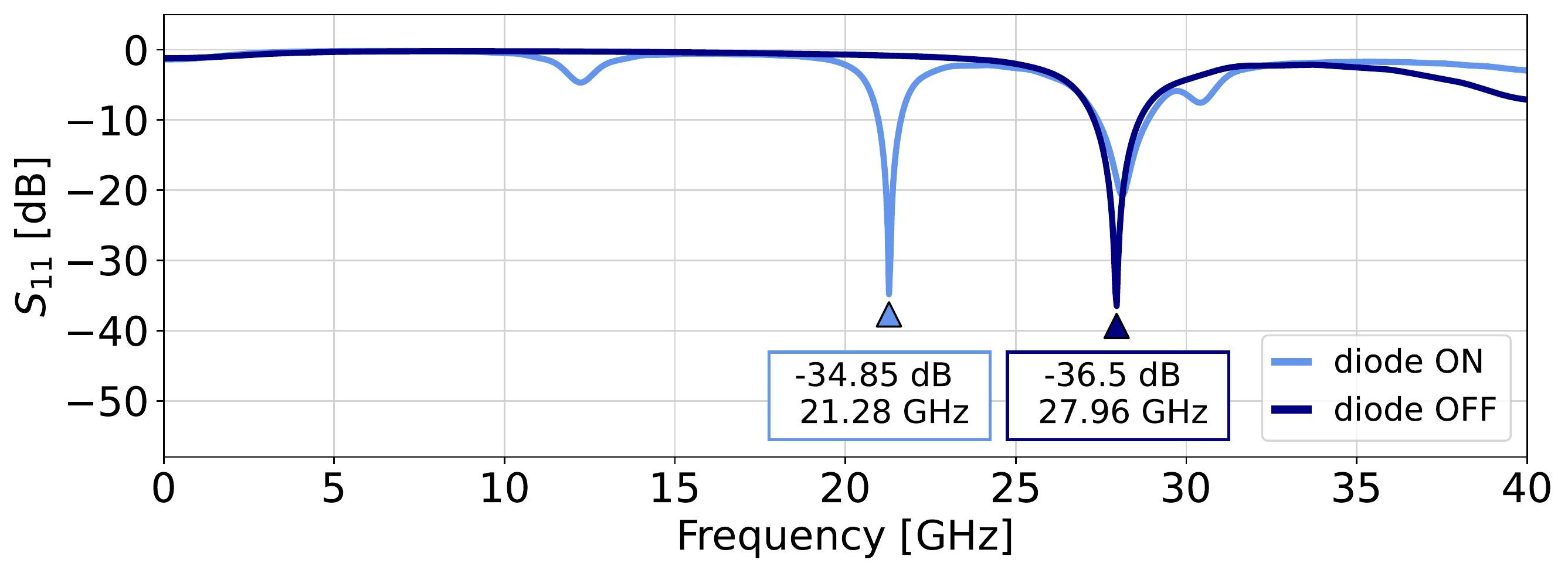}
    \caption{Full-wave simulations in CST of a single \gls{ris} element in terms of $S_{11}$ against the operating frequency for both diode states.}
    \label{fig:s11_comparison}
\end{figure}

\textbf{\gls{ris} element validation.} First, we feed a single \gls{ris} element with a transmission line and evaluate its $S_{11}$~\cite{balanis2012} when working either at $f_1$ or $f_2$, namely with diode \textsc{off} or with diode \textsc{on}, respectively. As shown in Fig.~\ref{fig:s11_comparison}, the $S_{11}$ strongly dips at $f_1$ (diode \textsc{off}) and at $f_2$ (diode \textsc{on}), reaching its global minimum in both cases. Besides, although the $S_{11}$ obtained with diode \textsc{on} shows additional local minima, their values are considerably higher ($\sim 15$ dB) than the global minimum at $f_2$, thus confirming the effectiveness of our design.

\textbf{\gls{ris} array response validation.} We evaluate now the 2D array response of an \gls{ris} with $N_x = N_y = 10$ elements, with unit cells given as in Fig.~\ref{fig:patch_antenna}. Since it is not possible to reduce the effective inter-element distance, we set $d=0.56\lambda_1$, where $\lambda_1$ corresponds to the working frequency $f_1$. Hence, when operating at $f_2 < f_1$, the ratio between the corresponding wavelength and the inter-element distance $d/\lambda_2$ is lower than $0.5$, thus resulting in mutual coupling among the elements.
We would like to point out that by designing the inter-element distance based on $f_2$, \gls{ris} operation at $f_1 > f_2$ would likely result in a beampattern affected by grating lobes as the latter appear in array beampatterns whenever the inter-element distance is higher than a full working wavelength\footnote{Nonetheless, such design might be helpful in scenarios with sparse \gls{ue} distributions to effectively support multicast communication.}. 

\begin{figure}[t!]
    \centering  
    \includegraphics[clip,width =  \linewidth]{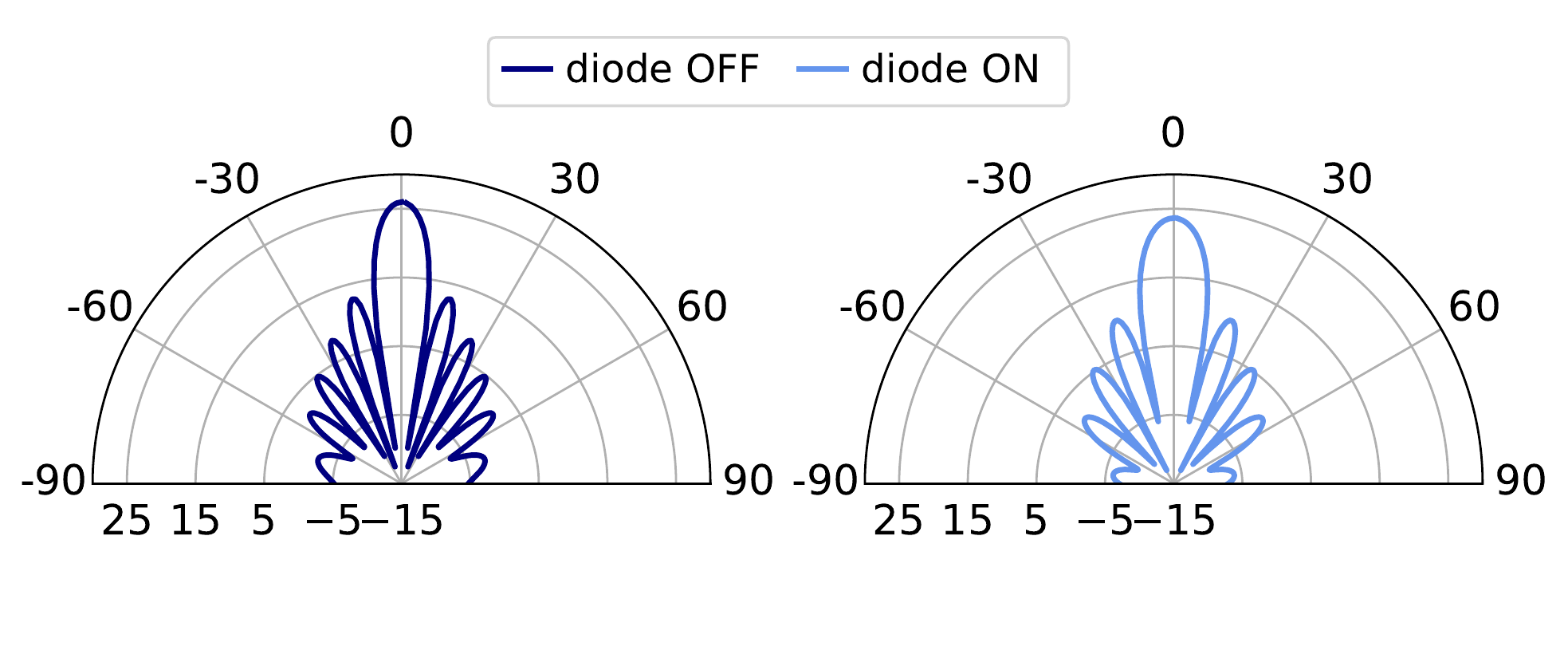}
    \caption{Beampattern obtained via full-wave simulations in CST of a $10 \times 10$ \gls{ris} with unit cells given as in Fig.~\ref{fig:patch_antenna}.}
    \label{fig:beampattern_comparison}
\end{figure}

Fig.~\ref{fig:beampattern_comparison} shows the \gls{ris} beampattern obtained via CST full-wave simulations for the two different diode states. In particular, on the left-hand side the \gls{ris} is operated at frequency $f_1$ and exhibits a narrow beampattern thanks to the large number of available antennas and the inter-element distance of $0.56\lambda_1$. Whereas, on the right-hand side the \gls{ris} is operated at frequency $f_2$, which results in a ratio of inter-element spacing over wavelength of 0.42. The obtained beampattern is characterized by a comparable main lobe magnitude with respect to the previous case and, as expected, the side lobes and the beamwidth appear to be increased.

\section{Numerical results}
\label{s:performance}

We present numerical results to validate our \name{} design by examining the performance of the multi-frequency \gls{ris} described in Section~\ref{s:design}, which comprises of $N_x=N_y = 10$ elements and is designed to have an inter-element distance of $d=0.56 \lambda_1$, with $\lambda_1$ corresponding to the maximum supported working frequency of $f_1 = 27.96$ GHz. While such design behaves as a conventional \gls{ris} at the working frequency $f_1$, in this section we focus on the effect of mutual coupling on the array response and the benefits of the procedure described in Section~\ref{s:algorithm} that aims at mitigating such unwanted effects. Hence, we set the working frequency to $f=f_2$, with $f_2 = 21.28$ GHz, which corresponds to an inter-element spacing over wavelength ratio of $0.42$.

\begin{figure}[t!]
    \centering  
    \includegraphics[clip,width =  \linewidth]{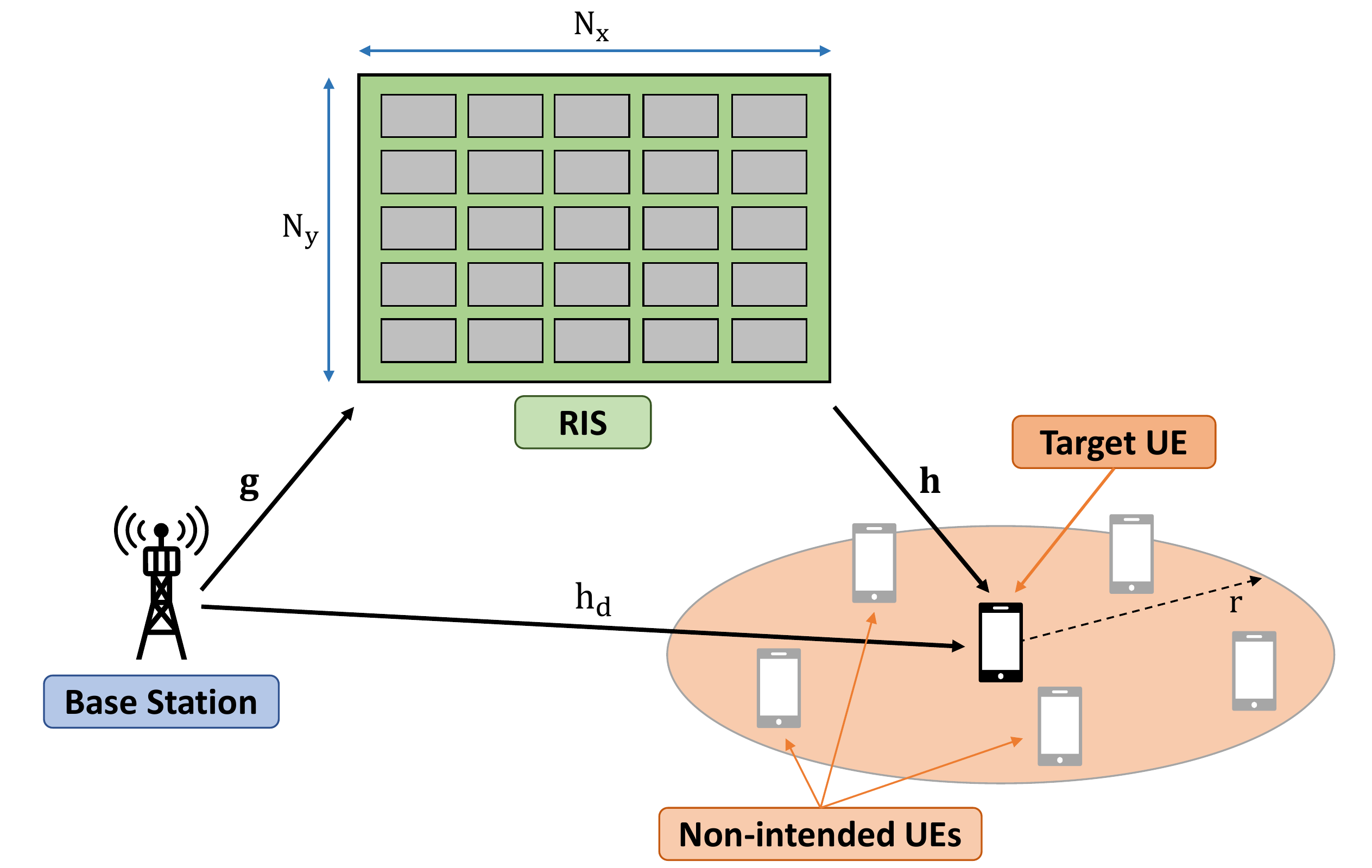}
    \caption{Considered simulation scenario}
    \label{fig:scenario}
\end{figure}

The considered scenario is depicted in Fig.~\ref{fig:scenario}, where the \gls{ris} is placed in position $\vv{p}_{\rmris} = \begin{bmatrix}10 & 20 & 0\end{bmatrix}^\tran$, whereas the \gls{ue} is placed at $\vv{p}_{\rmue} = \begin{bmatrix}10 & 0 & -10\end{bmatrix}^\tran$. Furthermore, we set $\sigma_n^2 = -80$ dBm and $P = 24$ dBm. We assume that there are $T=20$ non-intended \glspl{ue} uniformly distributed over a circle of (variable) radius $r$ around the target \gls{ue} and we average our results over $10^3$ Monte Carlo independent realizations of their positions. We compare the performance of \name{} against a \emph{naive} approach where all the \gls{ris} elements are activated, with no concern for the effect in terms of mutual coupling on the resulting array response. 

\begin{figure}[t]
    \centering  
    \includegraphics[clip,width =  \linewidth]{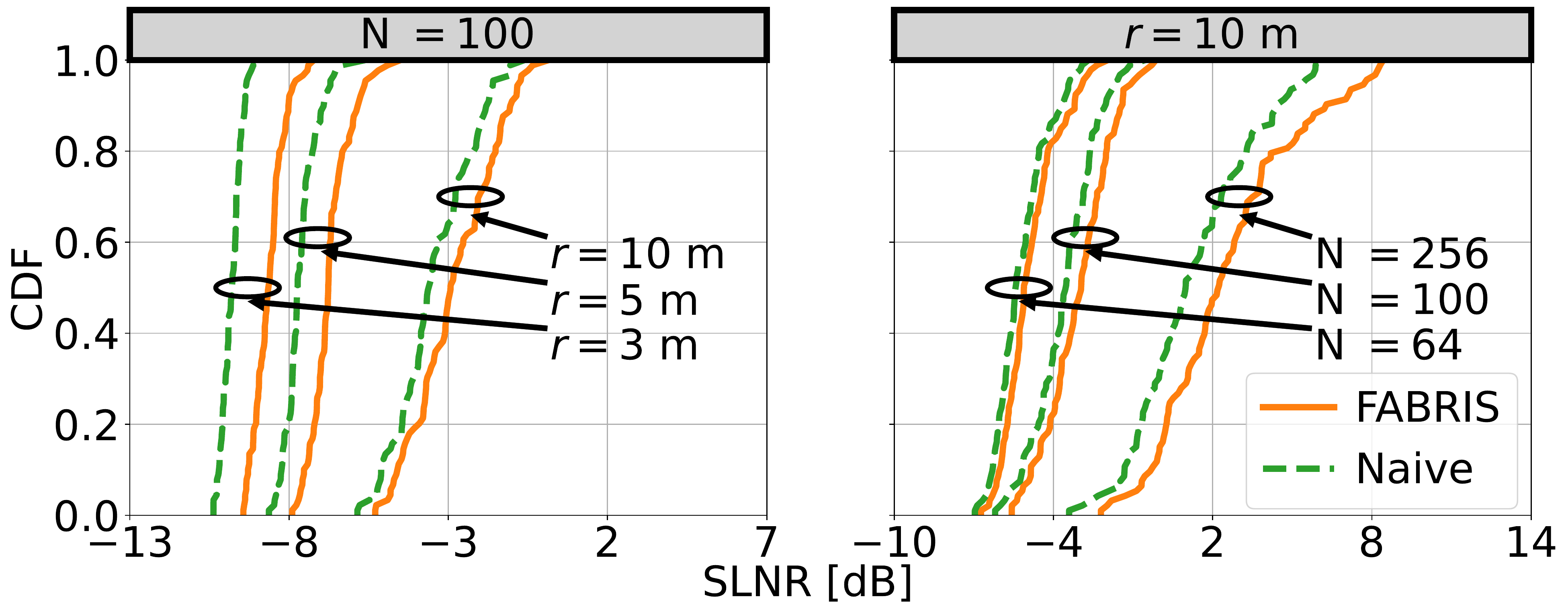}
    \caption{\Gls{cdf} of the \gls{slnr} at the target \gls{ue} obtained with both the proposed \name{} and the reference naive approaches for different values of $r$ with $N=100$ (left-hand side), and for different values of $N$ with $r=10$ m (right-hand side).}
    \label{fig:CDF}
\end{figure}

Fig.~\ref{fig:CDF} shows on the left-hand side the \gls{cdf} of the \gls{slnr} obtained with both the proposed \name{} method and with the reference naive approach for different values of the radius $r$ over which the non-intended \glspl{ue} are scattered. We notice that \name{} effectively reduces leakage to neighbouring \glspl{ue} by suitably optimizing the \gls{ris} activation profile as to mitigate the effects of mutual coupling, without excessively compromising beamforming gain, thus obtaining a \emph{sweet spot} in the trade-off between the two. This effect is more evident for smaller values of $r$, since the non-intended \glspl{ue} may be very close to the target \gls{ue} and thus experience strong leakage.

\medskip 

\hspace{-0.4cm}\begin{minipage}[t]{\linewidth}
  \begin{minipage}[b]{0.49\linewidth}
    \centering  
    \includegraphics[clip,width =  \linewidth]{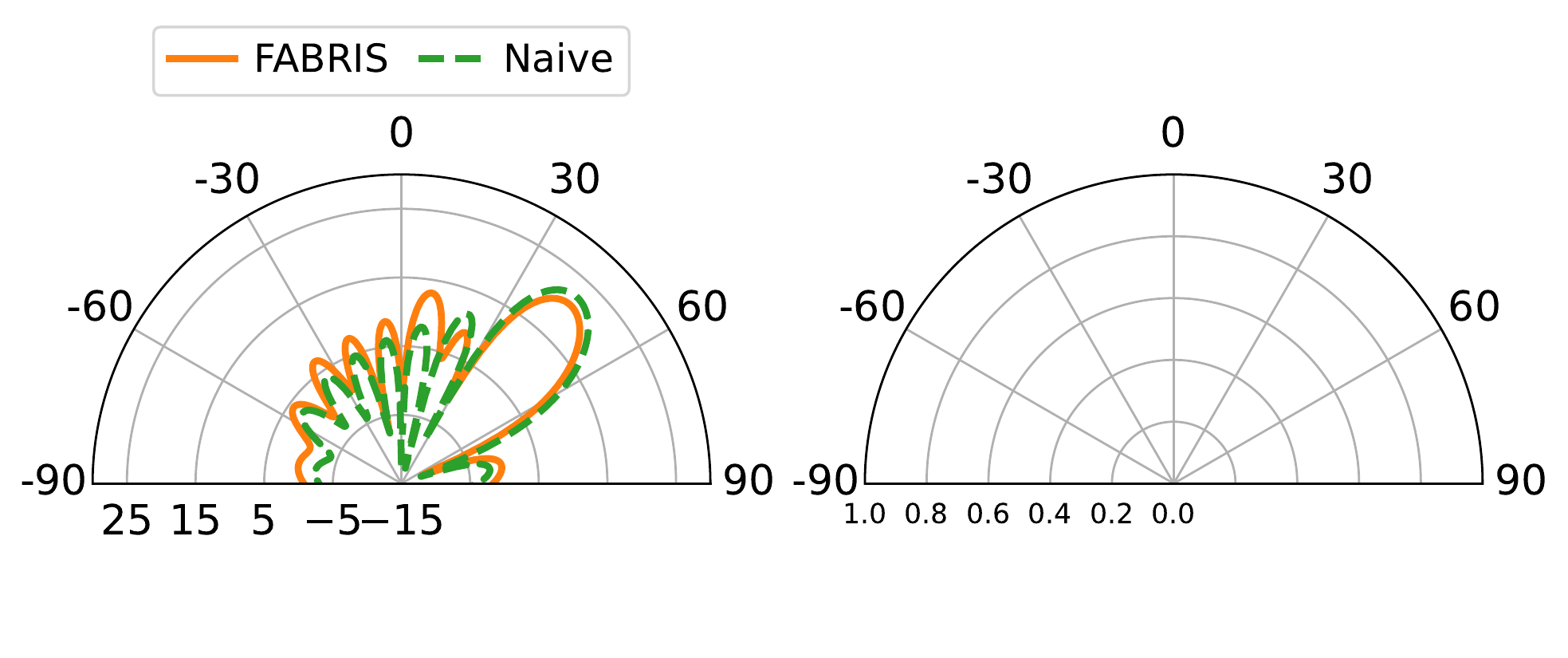}
    \vspace{-0.5cm}
  \end{minipage}
  \hfill
  \begin{minipage}[b]{0.49\linewidth}
    \centering
    \resizebox{.9\linewidth}{!}{%
    \begin{tabular}{c|c|c}
                                                                                & Naive & \name{} \\ \hline 
     \rowcolor[HTML]{EFEFEF}
    \begin{tabular}[c]{@{}c@{}}Main lobe \\ magnitude {[}dBi{]}\end{tabular}    & 22.3   & 20.8  \\ 
    \begin{tabular}[c]{@{}c@{}}Angular width \\ (3dB) {[}degree{]}\end{tabular} & 16.0   &  15.4 \\ 
    \rowcolor[HTML]{EFEFEF}
    \begin{tabular}[c]{@{}c@{}}Side lobe \\ level {[}dB{]}\end{tabular} & -10.8 &  -7.7 \\ 
    \end{tabular}
    }
    \vspace{0.3cm}
    \end{minipage}
    \captionof{figure}{CST full-wave simulation of the beampattern for a $10 \times 10$ \gls{ris} obtained by \name{} against the reference naive approach, with $r=10$ m and $N=100$. Relevant properties are summarized in the right-hand-side table.}
    \label{fig:beampattern_comparison_FABRIS_naive}
\end{minipage} 

\medskip

The right-hand side of Fig.~\ref{fig:CDF} shows the \gls{cdf} of th \gls{slnr} for both considered approaches in the case of $r=10$ m and for different values of $N$\footnote{For simplicity, we assume to have a squared \gls{ris}, i.e., $N_x = N_y$, although \name{} may be readily applied to any rectangular \gls{ris} design.}. Here, a lower value of $N$ leads to almost identical performances, since there are only few degrees of freedom to exploit in order to improve the directivity of the effective array response at the \gls{ris}. Whereas, for larger values of $N$ the proposed \name{} method obtains highly-selective beamforming at a small cost in terms of beamforming gain at the target \gls{ue}.

As an illustrative example, Fig.~\ref{fig:beampattern_comparison_FABRIS_naive} shows the full-wave simulation in CST of the antenna array response at the \gls{ris} obtained with both the proposed \name{} approach and with the reference naive method for a single realization in the case of $N=100$ and $r=5$ m. The proposed \name{} approach reduces the main lobe angular width of $0.6^\circ$ at the expense of a small penalty in terms of main lobe magnitude. Besides, although the side lobe level (mainly provided by the beam pointing at $\sim 10^\circ$) appears to be increased by $3.1$~dB, the leakage level in the vicinity of the \gls{ue}, i.e. in the directions within the circle of radius $r = 5$ m centered around the \gls{ue}, is greatly reduced, thereby improving the selected \gls{slnr} metric.  

Lastly, we would like to remark that while in this paper we make a first attempt at designing a multi-frequency \gls{ris}, we expect much larger gains brought by the proposed \name{} approach when moving from binary to higher-order and wider ranges of supported frequencies, which is left as future work.

\section{Conclusions}

In this paper, we presented \name{}, a practical \gls{ris} design enabling operation at multiple frequencies. \name{} builds upon a novel patch antenna layout involving two sub-elements connected via a \gls{pin} diode whose bias can be controlled to set the \textsc{on} or \textsc{off} states. The proposed hardware design is complemented by a novel optimization approach that outputs the \gls{ris} both the phase shifts configuration, which is obtained via microstrip-lines of different lengths, and the sub-elements activation profile as to reduce signal leakage in unwanted directions. Our results, involving synthetic and full-wave simulations, show that optimizing both variables leads to substantial performance gains with respect to a reference naive approach that does not consider the leakage generated by multi-frequency antenna arrays. Future work will consider the case of multiple antennas  and both higher-order and wider ranges of supported operating frequencies.   



\bibliographystyle{IEEEtran}
\bibliography{IEEEabrv,bibliography}

\begin{thebibliography}{10}
\providecommand{\url}[1]{#1}
\csname url@samestyle\endcsname
\providecommand{\newblock}{\relax}
\providecommand{\bibinfo}[2]{#2}
\providecommand{\BIBentrySTDinterwordspacing}{\spaceskip=0pt\relax}
\providecommand{\BIBentryALTinterwordstretchfactor}{4}
\providecommand{\BIBentryALTinterwordspacing}{\spaceskip=\fontdimen2\font plus
\BIBentryALTinterwordstretchfactor\fontdimen3\font minus
  \fontdimen4\font\relax}
\providecommand{\BIBforeignlanguage}[2]{{%
\expandafter\ifx\csname l@#1\endcsname\relax
\typeout{** WARNING: IEEEtran.bst: No hyphenation pattern has been}%
\typeout{** loaded for the language `#1'. Using the pattern for}%
\typeout{** the default language instead.}%
\else
\language=\csname l@#1\endcsname
\fi
#2}}
\providecommand{\BIBdecl}{\relax}
\BIBdecl

\bibitem{direnzo2020}
M.~Di~Renzo, A.~Zappone, M.~Debbah, M.-S. Alouini, C.~Yuen, J.~de~Rosny, and
  S.~Tretyakov, ``{Smart Radio Environments Empowered by Reconfigurable
  Intelligent Surfaces: How It Works, State of Research, and The Road Ahead},''
  \emph{IEEE Journal on Selected Areas in Communications}, vol.~38, no.~11, pp.
  2450--2525, 2020.

\bibitem{Dai2020}
L.~Dai, M.~D. Renzo, C.~B. Chae, L.~Hanzo, B.~Wang, M.~Wang, X.~Yang, J.~Tan,
  S.~Bi, S.~Xu, F.~Yang, and Z.~Chen, ``{Reconfigurable Intelligent
  Surface-Based Wireless Communications: Antenna Design, Prototyping, and
  Experimental Results},'' \emph{IEEE Access}, vol.~8, pp. 45\,913--45\,923,
  2020.

\bibitem{Mursia2020}
P.~Mursia, V.~Sciancalepore, A.~Garcia-Saavedra, L.~Cottatellucci,
  X.~Costa-Perez, and D.~Gesbert, ``{RISMA: Reconfigurable Intelligent Surfaces
  Enabling Beamforming for IoT Massive Access},'' \emph{IEEE Journal on
  Selected Areas in Communications}, vol.~39, no.~4, pp. 1072--1085, 2021.

\bibitem{albanese2021}
A.~Albanese, V.~Sciancalepore, and X.~Costa-Pérez, ``{First Responders Got
  Wings: UAVs to the Rescue of Localization Operations in Beyond 5G Systems},''
  \emph{IEEE Communications Magazine}, vol.~59, no.~11, pp. 28--34, 2021.

\bibitem{albanese2022}
A.~Albanese, F.~Devoti, V.~Sciancalepore, M.~Di~Renzo, and X.~Costa-P\'erez,
  ``{MARISA: A Self-configuring Metasurfaces Absorption and Reflection Solution
  Towards 6G},'' in \emph{IEEE INFOCOM 2022 - IEEE Conference on Computer
  Communications}, 2022.

\bibitem{RIScommag_2021}
{E. Calvanese-Strinati, et al.}, ``Reconfigurable, intelligent, and sustainable
  wireless environments for 6g smart connectivity,'' \emph{IEEE Communications
  Magazine}, vol.~59, no.~10, pp. 99--105, 2021.

\bibitem{Trichopoulos2021}
\BIBentryALTinterwordspacing
G.~C. Trichopoulos, P.~Theofanopoulos, B.~Kashyap, A.~Shekhawat, A.~Modi,
  T.~Osman, S.~Kumar, A.~Sengar, A.~Chang, and A.~Alkhateeb, ``{Design and
  Evaluation of Reconfigurable Intelligent Surfaces in Real-World
  Environment},'' 2021. [Online]. Available:
  \url{http://arxiv.org/abs/2109.07763}
\BIBentrySTDinterwordspacing

\bibitem{Gros2021}
J.-B. Gros, V.~Popov, M.~A. Odit, V.~Lenets, and G.~Lerosey, ``{A
  Reconfigurable Intelligent Surface at mmWave Based on a Binary Phase Tunable
  Metasurface},'' \emph{IEEE Open Journal of the Communications Society},
  vol.~2, pp. 1055--1064, 2021.

\bibitem{Qingqing2021}
Q.~Wu, S.~Zhang, B.~Zheng, C.~You, and R.~Zhang, ``{Intelligent Reflecting
  Surface-Aided Wireless Communications: A Tutorial},'' \emph{IEEE Transactions
  on Communications}, vol.~69, no.~5, pp. 3313--3351, 2021.

\bibitem{Dunna2020}
M.~Dunna, C.~Zhang, D.~Sievenpiper, and D.~Bharadia, ``{ScatterMIMO: enabling
  virtual MIMO with smart surfaces},'' in \emph{Proceedings of the 26th Annual
  International Conference on Mobile Computing and Networking (MOBICOM)}, 2020.

\bibitem{Pei2021}
X.~Pei, H.~Yin, L.~Tan, L.~Cao, Z.~Li, K.~Wang, K.~Zhang, and E.~Björnson,
  ``{RIS-Aided Wireless Communications: Prototyping, Adaptive Beamforming, and
  Indoor/Outdoor Field Trials},'' \emph{IEEE Transactions on Communications},
  vol.~69, no.~12, pp. 8627--8640, 2021.

\bibitem{Aftab2015}
N.~Aftab, H.~T. Chattha, Y.~Jamal, A.~Sharif, and Y.~Huang, ``{Reconfigurable
  patch antenna for wireless applications},'' in \emph{Proceedings of the 9th
  European Conference on Antennas and Propagation (EuCAP)}, 2015.

\bibitem{balanis2012}
C.~Balanis, \emph{{Antenna Theory: Analysis and Design}}.\hskip 1em plus 0.5em
  minus 0.4em\relax Wiley, 2012.

\bibitem{CST}
\BIBentryALTinterwordspacing
{Dassault Systèmes Simulia Corp.}, ``{CST Studio Suite 2021}.'' [Online].
  Available:
  \url{https://www.3ds.com/products-services/simulia/products/cst-studio-suite/}
\BIBentrySTDinterwordspacing

\bibitem{Kishk_13}
A.~Kishk, \emph{{Advancement in Microstrip Antennas with Recent
  Applications}}.\hskip 1em plus 0.5em minus 0.4em\relax IntechOpen, 2013.

\bibitem{pin}
\BIBentryALTinterwordspacing
\emph{PIN Diode Basics. Application Note.}, Skyworks Solutions, Inc., 2008.
  [Online]. Available:
  \url{https://www.skyworksinc.com/-/media/SkyWorks/Documents/Products/1-100/200823A.pdf}
\BIBentrySTDinterwordspacing

\end{thebibliography}

\end{document}